# Real-World Recommender Systems for Academia: The Pain and Gain in Building, Operating, and Researching them [Long Version][1]

Joeran Beel[1,2] and Siddharth Dinesh[3]

[1]Trinity College Dublin, Department of Computer Science, ADAPT Centre, Ireland
joeran.beel@adaptcentre.ie
[2]National Institute of Informatics, Digital Content and Media Sciences Research Division, Tokyo, Japan
beel@nii.ac.jp
[3]Birla Institute of Technology and Science, Pilani, India
f2012519@goa.bits-pilani.ac.in

**Abstract.** Research on recommender systems is a challenging task, as is building and operating such systems. Major challenges include non-reproducible research results, dealing with noisy data, and answering many questions such as how many recommendations to display, how often, and, of course, how to generate recommendations most effectively. In the past six years, we built three research-article recommender systems for digital libraries and reference managers, and conducted research on these systems. In this paper, we share some experiences we made during that time. Among others, we discuss the required skills to build recommender systems, and why the literature provides little help in identifying promising recommendation approaches. We explain the challenge in creating a randomization engine to run A/B tests, and how low data quality impacts the calculation of bibliometrics. We further discuss why several of our experiments delivered disappointing results, and provide statistics on how many researchers showed interest in our recommendation dataset.

## 1 Introduction

Recommender systems is a fascinating topic for both researchers and industry. Researchers find in recommender systems a topic that is relevant for many disciplines: machine learning, text mining, artificial intelligence, network analysis, bibliometrics, databases, cloud computing, scalability, data science, visualization, human computer interaction, and many more. That makes research in recommender systems interesting and creates many opportunities to cooperate with other researchers. The premier conference in the field (ACM RecSys) has been regularly sold out in the past years (Figure 1), which demonstrates the growing interest in recommender systems. For industry, recommender systems offer an opportunity to provide additional value to customers by helping them finding relevant items. Recommender systems may also provide a justification to store user-related data, which may be used for generating additional revenue. In addition, recommender systems may even become a major part of the business model, as companies such as Netflix, Spotify, and Amazon demonstrate.

[1] This article is a long version of the article published in the Proceedings of the 5th International Workshop on Bibliometric-enhanced Information Retrieval (BIR) (Beel & Dinesh, 2017). This article contains all information from the original article plus additional details.

Over the past six years, we built, operated, and researched three research-article recommender systems in the context of digital libraries and reference management. The work was often rewarding, but also challenging and occasionally even painful. We share some of our experiences in this article. This article is not a research article but a mixture of a project report, lessons learned, text-book, and summary of our previous research, enriched with some novel research results.[2]

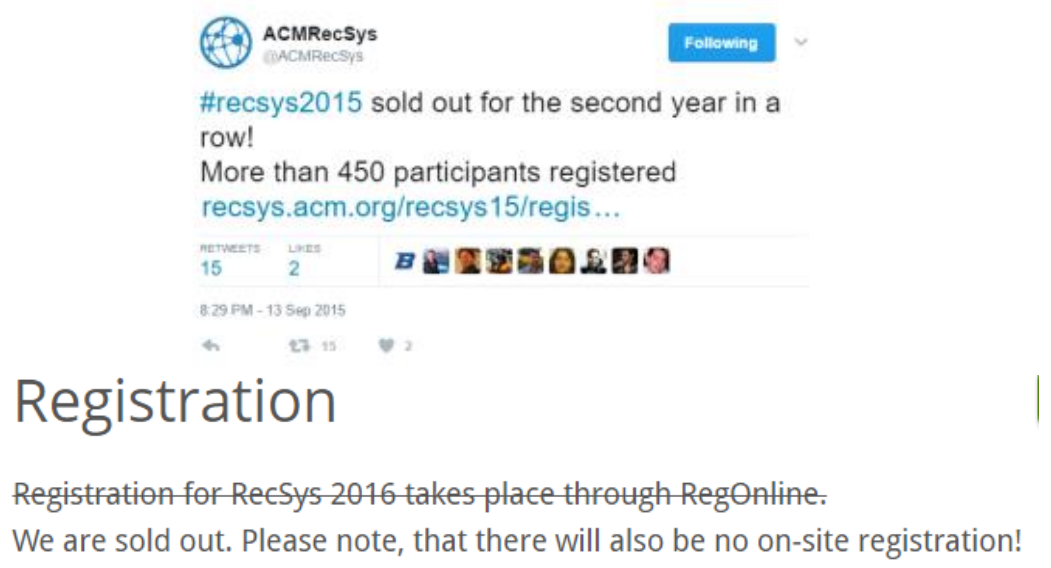

*Figure 1: RecSys Conference 2014, 2015, and 2016 were sold out[3,4]*

The primary audience of this article are researchers and developers who think about developing a real-world recommender system for research purposes, or for integrating the recommender system on-top of a real product. While our focus lies on recommender system in the context of digital libraries and reference managers, researchers and developers from other disciplines may also find some relevant information in our article. As this paper is an invited paper for the "5th International Workshop on Bibliometric-enhanced Information Retrieval", we particularly discuss our work in the context of bibliometric-enhanced recommender systems and information retrieval.

## 2 Our Recommender Systems

### 2.1 SciPlore MindMapping

In 2009, we introduced SciPlore MindMapping (Beel, Gipp, & Mueller, 2009). The software enabled researchers to manage their references, annotations, and PDF files in

[2] The data and scripts we used for the novel analyses is available at http://data.mr-dlib.org

[3] https://recsys.acm.org/recsys16/registration/

[4] https://twitter.com/acmrecsys/status/643129410233729024

mind-maps (Figure 2). In these mind-maps, users could create categories that reflect their research interests or that represent sections of a new manuscript. Users could then sort their PDF files, annotations, and references in these categories. In 2011, we integrated a recommender system in SciPlore MindMapping (Beel, 2011). The system was rather simple. Whenever users selected a node in a mind-map, the text of the node was sent as search query to Google Scholar, and the first three results of Google Scholar were shown as recommendations. For more details on SciPlore MindMapping and its recommender system, please refer to Beel, Langer, Kapitsaki, Breitinger, & Gipp (2015), Beel & Langer (2011) and Beel, Langer, Genzmehr, & Gipp (2014)

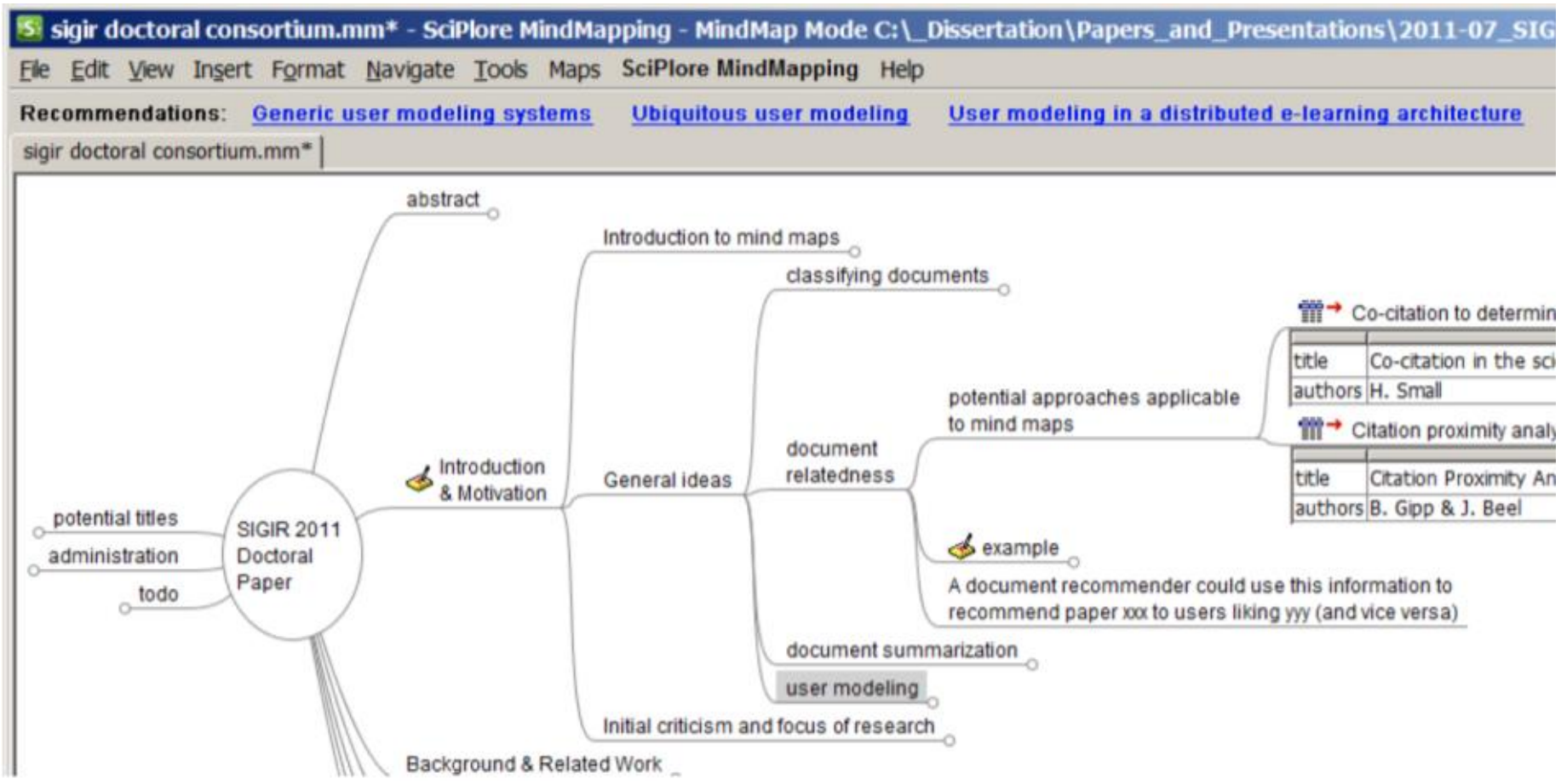

*Figure 2: Screenshot of SciPlore MindMapping*

## 2.2 Docear

Docear[5] is the successor of SciPlore MindMapping, pursuing the same goal, i.e. enabling researchers to manage their references, annotations, and PDF files in mind maps (Beel, Gipp, Langer, & Genzmehr, 2011).[6] In contrast to SciPlore MindMapping, Docear has more features, a neater interface (Figure 3), and a more sophisticated recommender system (Beel, Langer, Genzmehr, & Nürnberger, 2013a; Beel, Langer, Gipp, & Nürnberger, 2014). The recommender system features a comprehensive user modeling engine, and uses Apache Lucene/Solr for content-based filtering as well as some proprietary implementations of other recommendation approaches. Docear is a desktop software but transfers users' mind maps to Docear's servers. On the servers, Docear's recommender system calculates user specific recommendations. Recommendations are shown every couple of days to users (Figure 4), and users may also request recommendations explicitly. Docear has a corpus of around 2 million full-text documents freely available on the Web.

[5] http://docear.org

[6] Currently, Docear’s recommender system is offline because we focus on the development of Mr. DLib.

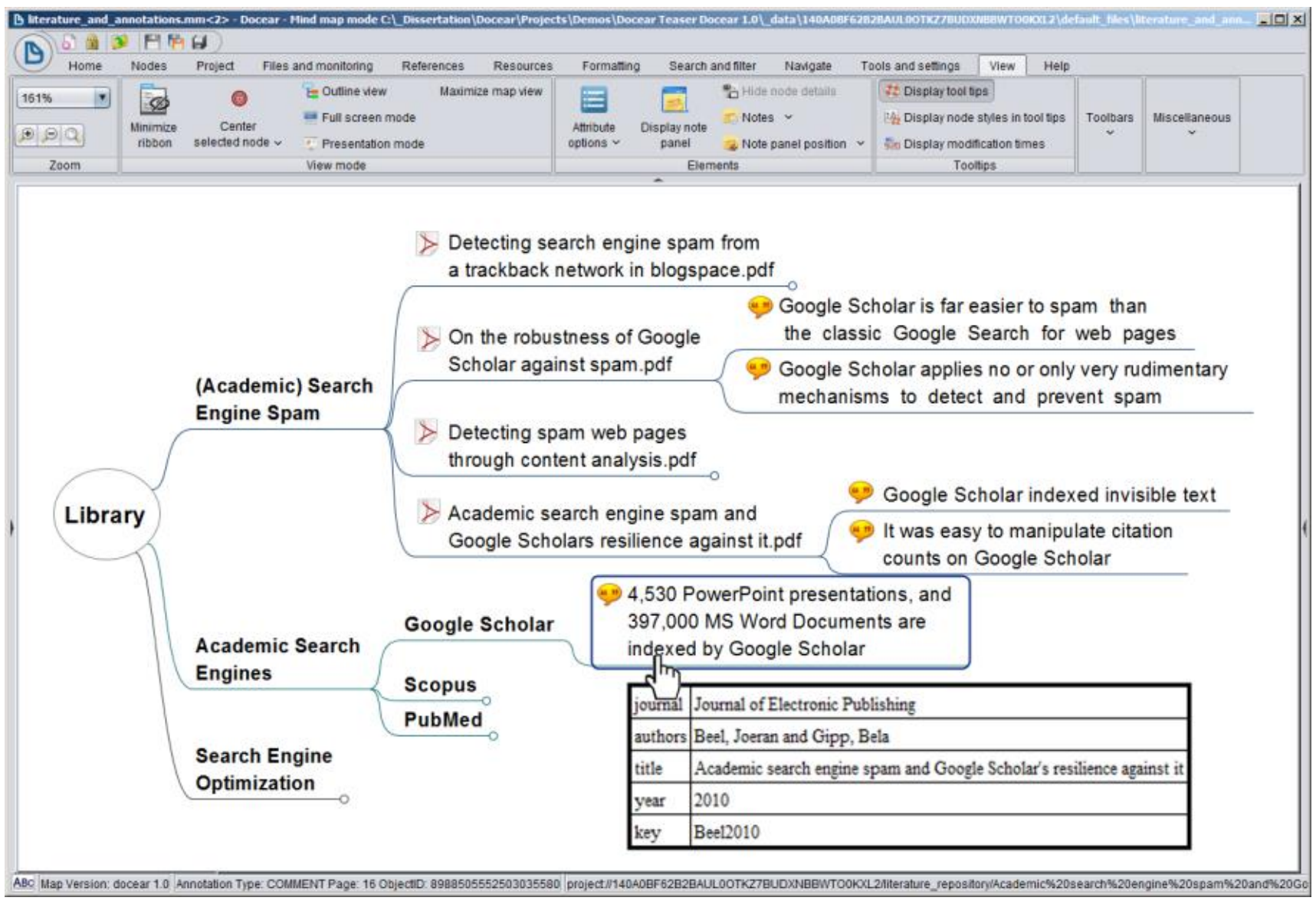

*Figure 3: Screenshot of Docear*

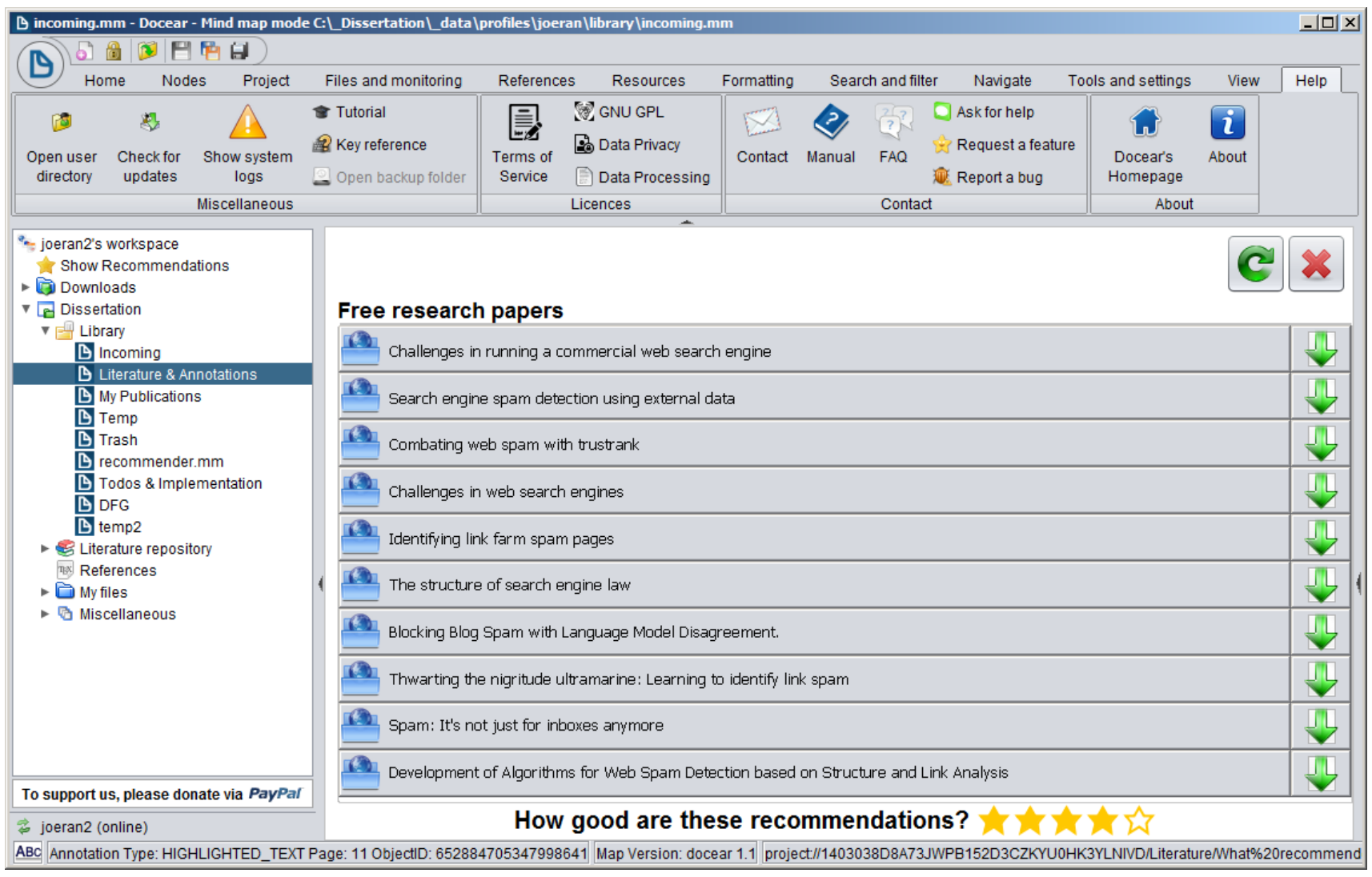

*Figure 4: Screenshot of recommendations in Docear*

## 2.3 Mr. DLib

Our latest recommender system is Mr. DLib[7], a machine-readable digital library that provides recommendations as-a-service to third parties (Beel, Gipp, & Aizawa, 2017). This means, third parties such as such as digital libraries and reference managers can easily integrate a recommender system into their product via Mr. DLib. The recommender system is hosted and operated by Mr. DLib and partners only need to request recommendations for a specific item via a REST API (Figure 5).

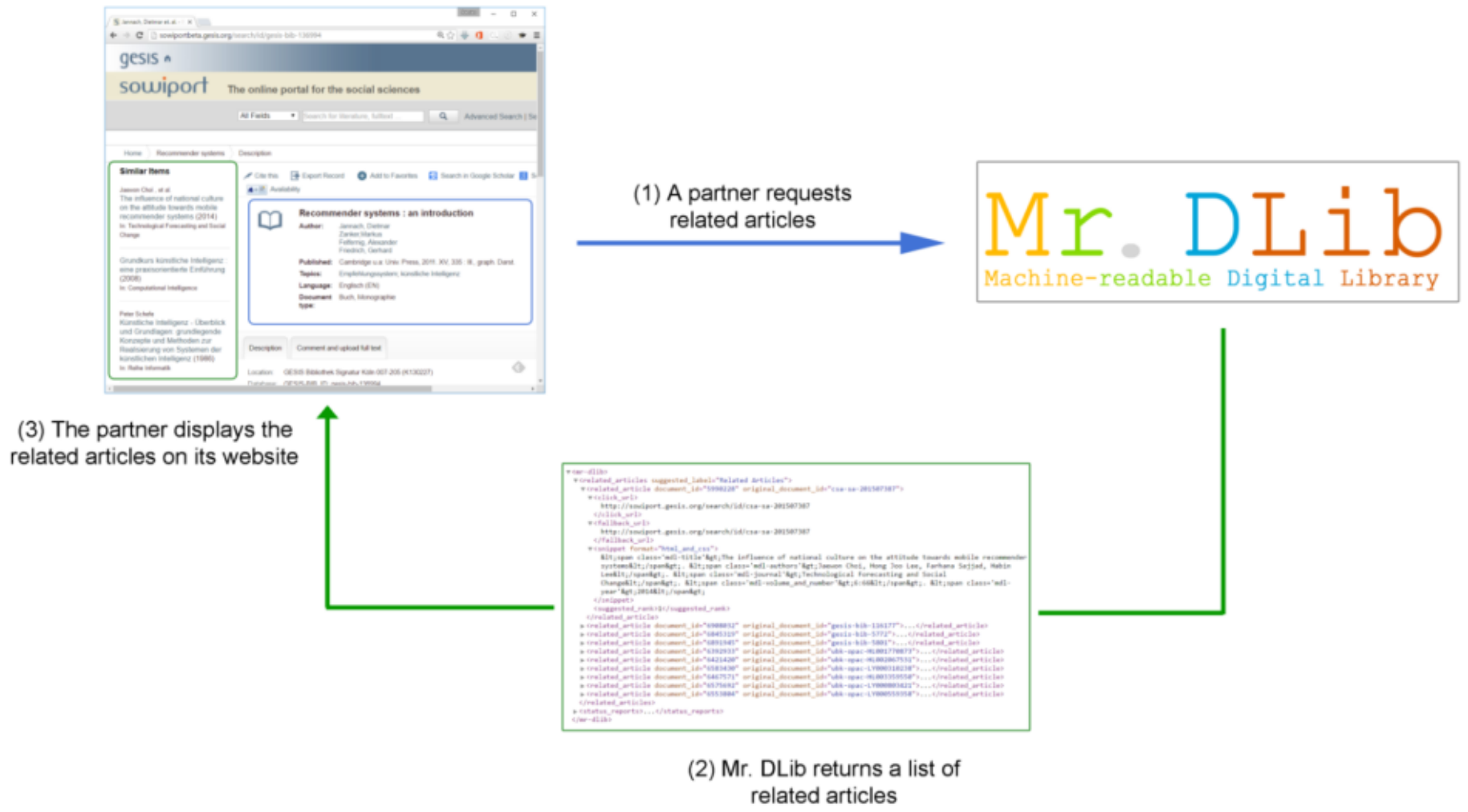

*Figure 5: Illustration of Mr. DLib's recommendation process*

Our first pilot partner is Sowiport (Hienert, Sawitzki, & Mayr, 2015). Sowiport is operated by 'GESIS – Leibniz-Institute for the Social Sciences', which is the largest infrastructure institution for the Social Sciences in Germany. Sowiport contains about 9.6 million literature references and 50,000 research projects from 18 different databases, mostly relating to the social and political sciences. Literature references usually cover keywords, classifications, author(s) and journal or conference information and if available: citations, references and links to full texts. We additionally integrated Mr. DLib into the reference manager JabRef (Feyer, Siebert, Gipp, Aizawa, & Beel, 2017), and currently discuss the integration with two more organizations.

Recommendations in Mr. DLib are primarily calculated with Apache Lucene/Solr, i.e. based on the metadata of the documents in Sowiports' corpus. We further experiment with stereotype and most-popular recommendations (Beel, Dinesh, Mayr, Carevic, & Raghvendra, 2017), as well as with enhanced content-based filtering based on key-phrase extraction and re-ranking with Mendeley readership statistics (Siebert, Dinesh, & Feyer, 2017). In the future, we plan to add further recommendation approaches; especially link/citation-based approaches seem promising (Schwarzer et al., 2016).

[7] http://mr-dlib.org

# 3 Recommender-System Development

## 3.1 The objective

In the research community, it is commonly assumed that the objective of a recommender system is to make users "happy" (Ge, Delgado-Battenfeld, & Jannach, 2010) by satisfying their “needs” (McNee, Kapoor, & Konstan, 2006). However, the objective of a recommender system operator might differ. Operators may also want to keep down costs for labor, disk storage, memory, CPU power, and traffic (Ricci, Rokach, Shapira, & Paul, 2011). Therefore, for operators, an effective recommender system may be one that can be developed, operated, and maintained at a low cost. Operators may also want to generate a profit from the recommender system (Gunawardana & Shani, 2009). Such operators might prefer to recommend items with higher profit margins, even if user satisfaction was not optimal. For instance, publishers might be more interested in recommending papers the user would have to pay for than papers the user could freely download. Before starting to develop a recommender system, one should think very careful about the recommender system’s objective – and how to measure if the objective was achieved (Beel & Langer, 2015).

## 3.2 The stakeholders

In a normal research environment, the number of stakeholders is limited. In a small research project, there are typically two or three people involved such as a Master student, a PhD student, a professor, or maybe a postdoctoral researcher. Even in somewhat bigger projects, stakeholders usually have a similar interest, i.e. producing research results that can be used for a publication, thesis, or grant proposal. In addition, the roles are relatively clear, i.e. there is a strict hierarchy between a student, postdoctoral researcher and a professor. However, with a real product, especially an “as-a-service” product, the number of stakeholders increases, and their interests may be quite different. Hence the need for communication, the duration until decisions are made, and the potential for conflicts increases.

For instance, within our research group, tasks are managed in a ticket system, source code is on GitHub, and large files are exchanged via Dropbox or OneDrive. However, project partners of Mr. DLib have different customs of managing tickets and source code, and exchanging files. The more partners Mr. DLib has, the more difficult the communication becomes. If we want to change something in Mr. DLib that affects multiple partners (e.g. logging some user-related data), we need to discuss this issue with each partner, and each one may have different opinions or goals. In the worst case, a partner says “we want it this way, or not at all”. To avoid such situations, it is crucial to clarify expectations before a project starts, and to be ready to not start a project with a certain partner. It should also be noted that partners from the industry just have different interests than researchers. A company typically wants to maximize profit and is usually not keen to publish technical details on how their products work. This attitude is not always compatible with the interests of researchers.

### 3.3 Required skills

To build and operate a recommender system, more than just knowledge about recommender systems and related disciplines such as text mining and machine learning is required. Server administration, databases, web technologies, data formats (e.g. XML or JSON), and data processing (crawling, parsing, transforming) are probably the most important ones, but also knowledge about software engineering in general (e.g. agile development, unit testing, etc.), scalability, data privacy laws, and project management is helpful. Niche knowledge that does not directly relate to recommender systems may also be beneficial and lead to novel recommendation approaches. For instance, knowledge in bibliometrics could help to develop novel re-ranking algorithms for content-based research-paper recommender systems (Cabanac, 2011). In such systems, a list of recommendation candidates, would be re-ranked based on e.g. how many citations the candidate papers have, or based on the h-index of the candidate papers' authors, which, however, then might strengthen the Mathew effect (Cabanac & Preuss, 2013). We have experimented with such algorithms but were only partly successful – maybe because we lack the expert-knowledge in bibliometrics (Siebert et al., 2017).

### 3.4 Required hardware

The literature provides little information about which hardware is required to host a recommender system. In the domain of digital libraries, we are aware only of some numbers that were reported by Jack (2012a) for Mendeley's recommender system. According to Jack, costs for hosting Mendeley's recommender system on Amazon S3 are 66$ per month plus $30 to update the recommender system. That system could deal with 20 requests per second generated by 20 million users.

SciPlore MindMapping's recommender system requires no additional hardware, because the user modeling was done locally on the users' computers, and the user model was then sent to Google Scholar.

Docear's recommender system runs on two servers. The first server is an Intel Core i7 PC with two 120GB SSDs, one 3 TB HDD, and 16 GB RAM. It runs a PDF Spider, PDF Analyzer, and the mind-map database, and its load is usually high, because web crawling and PDF processing require many resources. The second server is an Intel Core i7 with two 750 GB HDDs, and 8 GB RAM. It runs all other services including a Web Service to upload mind maps and request recommendations, a mind-map parser, MySQL database, Lucene, and an offline evaluator. The server load is rather low on average, which is important, because the Web Service is not only needed for recommendations but also for other tasks such as user registration. For more details, please refer to Beel, Langer, Gipp, et al. (2014).

Mr. DLib's recommender system uses three servers. The first server is a dedicated root server for 75€ per month and features an Intel i7-4790k with 32GB RAM, one 960GB SSD and one 2 TB HDD. The server is our development system and "work horse", i.e. we use it for resource intensive tasks such as parsing partners' content, requesting and processing readership data from Mendeley, and extracting key-phrases from the documents. The second server is a Virtual Private Server (VPS) with four cores, 14 GB RAM, and 1 TB HDD (with SSD boost). We use this server as Beta system for testing. The costs for renting this VPS is 12€ per month. The third server is the production system and similarly equipped as the development server: Intel i7-4790k with 32GB RAM, one 960GB SSD and one 1 TB HDD. Costs are 70€ per month.

### 3.5 Required labor

The labor that may be invested in building a recommender system is almost unlimited. We once had a meeting with a Netflix engineer and he told us that more than 100 software engineers would be working at Netflix only on the recommender system. Recently, Netflix changed a part in its recommendation algorithm that seemed rather minor – nevertheless, around 70 software engineers were involved in that process[8]. In addition, Netflix is paying 40 freelancers to manually add tags to movies[9]. In other words, Netflix is spending millions of Dollars every year on its recommender system, and even Netflix recommender system is far from perfect.

The recommender system for SciPlore MindMapping was developed mostly by one student within a few weeks. However, the system was very simple and cannot be considered a fully-fledged recommender system. For Docear, we – i.e. three full-time computer scientists supported occasionally by some students and volunteers – invested around 18 person-months within three years to implement the recommender system as it is today[10]. This estimate does not include the development for the Docear Desktop software.

For Mr. DLib, two students and a post-doctoral researcher spent 2 months almost full-time to develop a first prototype that was stable to enough to be used by our partner Sowiport. The development included setting up the server, designing the database, implement some simple logging mechanisms, import the data from Sowiport, design the web service, and support Sowiport in integrating Mr. DLib into their website. The integration of Mr. DLib in our second partner JabRef took around 5 months FTE. This time was longer than expected and had various reasons. The JabRef team had very high expectations regarding code quality, we had to do some re-design of our database and logging mechanisms, and there were just many other minor issues. In addition to those 2 plus 5 month, many months more have been spend for implementing additional recommendation approaches and enhancements, analyzing data, creating the website http://mr-dlib.org, and setting up ticket systems and WIKIs. Overall, around 30 months FTE have been spent to develop Mr. DLib in the past twelve months.

Most of the work was done by students who had little experience in recommender systems, and as much experience in software development as one can expect from a Master or PhD student. A professional software engineer with knowledge in recommender system, certainly would have done the work faster. However, implementing all the small things, as pointed out in the remainder of this paper, takes plenty of time, even for experiences developers. So, even an experienced software developer will need at least some months to build a recommender system that works reasonable well in a production system. Nevertheless, even such a system would be far from delivering perfect recommendations. Looking at Mr. DLib's recommendations quality, there are still plenty of options for optimization.

[8] http://www.theverge.com/2016/2/17/11030200/netflix-new-recommendation-system-global-regional

[9] https://www.wired.com/2013/08/qq_netflix-algorithm/

[10] This is a very rough estimate, as we did not keep track of the exact working hours for the recommender system.

An alternative to develop one's own recommender system is using a recommender-system as a service. There are several such services. Some services (e.g. Mr. DLib) focus on niche markets like digital libraries, others focus on generic recommendation approaches that can be applied in almost any domain. Microsoft Azure's Recommendation API is one of these services. Using such a service is significantly cheaper than developing one's own recommender system. For instance, Microsoft charges 75 US$[11] for 100,000 recommendation request per month[12].

We suggest to build ones' own recommender system, when the recommender system should be a unique selling proposition, i.e. a unique feature that should distinguish ones' product from the competitors' products. In addition, the recommender system should provide enough benefit that it is worth investing several months of work and a continuous effort for maintenance. In all other situations, it is probably more sensible to use a recommender system as-a-service.

For researchers who are interested in conducting research with real users, the "as-a-service" concept is generally interesting. When we developed Docear, we spent most of the time in developing the reference management functionality, giving support to users, doing marketing etc. For the recommender-system development and research, we had only little time. Now, with Mr. DLib's recommendations-as-a-service, it is the opposite: We spend most of the time for research and development and there is little overhead for customer support, and "normal" product development that does not relate to the research. Yet, we have the advantage of doing research with real users, and even users from different partners.

### 3.6 No help from the literature

There are hundreds of research articles about recommender systems in Academia (Figure 6), and probably thousands of articles about recommender systems in other domains. One might expect that such a large corpus of literature would provide advice on how to build a recommender system, and which recommendation approaches to use. Unfortunately, this is not the case, at least in the domain of research-paper recommender systems. The reasons are manifold. Many recommendation approaches were not evaluated at all, compared against too simple baselines, evaluated with too few users, or evaluated with highly tweaked datasets (Beel, Gipp, Langer, & Breitinger, 2016). Consequently, the meaningfulness of the results is questionable.

Even if evaluations were sound, recommendation effectiveness may vary a lot. In other words, only because a recommendation approach performed well in one scenario, does not mean it will perform well in another scenario. For instance, the TechLens team proposed and evaluated several content-based filtering (CBF) and collaborative filtering (CF) approaches for research-paper recommendations. In one experiment, CF and CBF performed similarly well (McNee et al., 2002). In other experiments, CBF outperformed CF (Dong, Tokarchuk, & Ma, 2009; McNee et al., 2002; Torres, McNee, Abel, Konstan, & Riedl, 2004), and in some more experiments CF outperformed CBF (Ekstrand et al., 2010; McNee et al., 2006; Torres et al., 2004). In other words: it remains speculative how CBF and CF would perform in a scenario that differs from one of those used in the existing evaluations.

[11] https://azure.microsoft.com/en-us/pricing/details/cognitive-services/recommendations/

[12] Of course, users need to invest some time to manage their data in Microsoft's cloud.

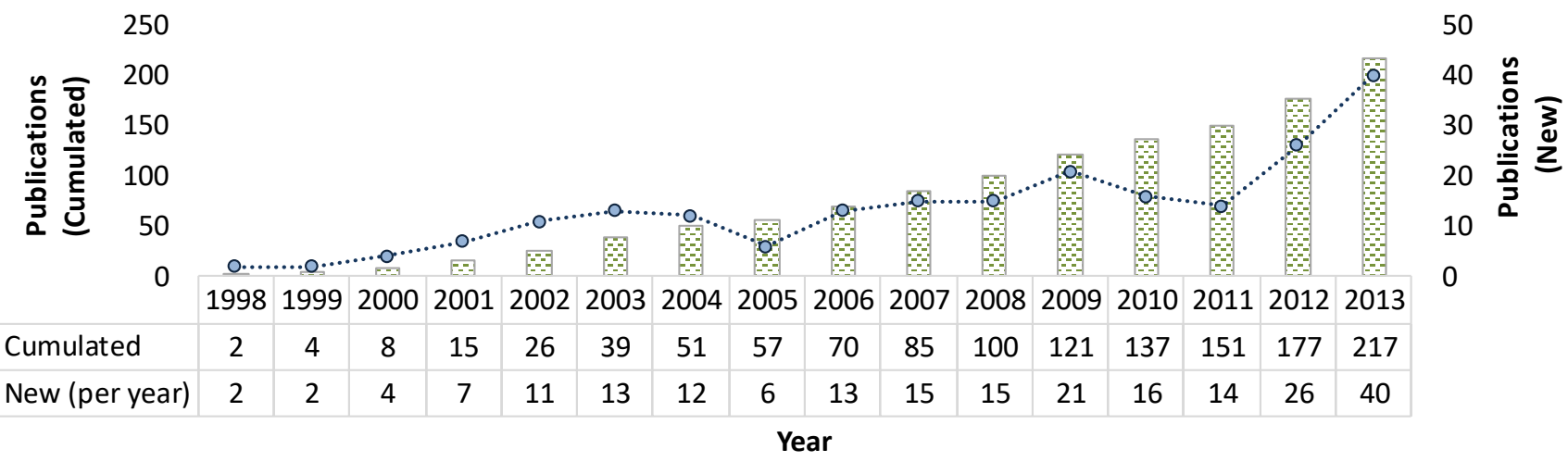

| | 1998 | 1999 | 2000 | 2001 | 2002 | 2003 | 2004 | 2005 | 2006 | 2007 | 2008 | 2009 | 2010 | 2011 | 2012 | 2013 |
|---|---|---|---|---|---|---|---|---|---|---|---|---|---|---|---|---|
| Cumulated | 2 | 4 | 8 | 15 | 26 | 39 | 51 | 57 | 70 | 85 | 100 | 121 | 137 | 151 | 177 | 217 |
| New (per year) | 2 | 2 | 4 | 7 | 11 | 13 | 12 | 6 | 13 | 15 | 15 | 21 | 16 | 14 | 26 | 40 |

*Figure 6: Number of published research articles about recommender systems in Academia* (Beel, Gipp, et al., 2016)

Some authors also used bibliometrics to enhance recommender systems in digital libraries. Popular metrics were PageRank (Bethard & Jurafsky, 2010), HITS (He, Pei, Kifer, Mitra, & Giles, 2010), Katz metric (He et al., 2010), citation counts (Bethard & Jurafsky, 2010; He et al., 2010; Rokach, Mitra, Kataria, Huang, & Giles, 2013), venues' citation counts (Bethard & Jurafsky, 2010; Rokach et al., 2013), citation counts of the authors' affiliations (Rokach et al., 2013), authors' citation count (Bethard & Jurafsky, 2010; Rokach et al., 2013), h-index (Bethard & Jurafsky, 2010), recency of articles (Bethard & Jurafsky, 2010), title length (Rokach et al., 2013), number of co-authors (Rokach et al., 2013), number of affiliations (Rokach et al., 2013), and venue type (Rokach et al., 2013). Again, results are not always coherent. For instance, Bethard and Jurafsky (2010) reported that using citation counts in the recommendation process *strongly* increased the effectiveness of their recommendation approach, while He et al. (2010) reported that citation counts *slightly* increased the effectiveness of their approach.

Our own research confirms that recommendation approaches perform very differently in different scenarios. We recently applied five recommendation approaches on six news websites (Beel, Breitinger, Langer, Lommatzsch, & Gipp, 2016). The results showed that recommendation approaches performing well on one news website performed poorly on others (Figure 7). For instance, the most-popular approach performed worst on tagesspiegel.de but best on cio.de.

There are several potential reasons for the unpredictability. In some cases, different evaluation methods were used. In other cases slight variations in algorithms or different user populations might have had an impact (Beel, Breitinger, et al., 2016; Beel, Langer, Nürnberger, & Genzmehr, 2013; Langer & Beel, 2014). However, it seems that, for instance, the operator of a news website cannot estimate how effective the most-popular recommendation approach would be until the operator has implemented and evaluated the approach on that particular website. Therefore, our advice is to read a recommender-system text book to get a general idea of recommender systems (Jannach, 2014; Konstan & Ekstrand, 2015; Ricci, Rokach, & Shapira, 2015). Then, choose a few recommendation frameworks and try to find the best recommendation approach for one's particular recommender system by implementing and evaluating the actual recommendation performance.

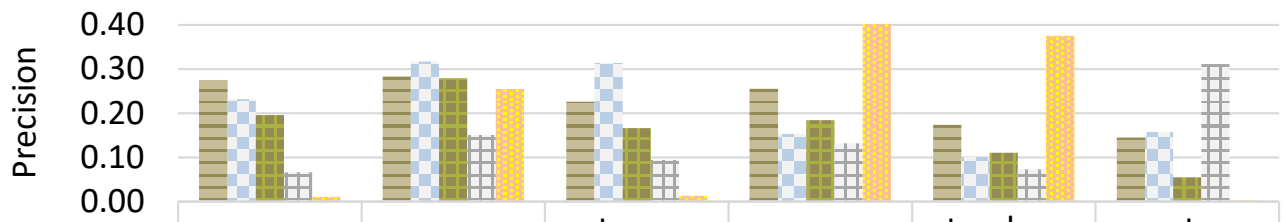

| | ksta.de | sport1.de | tages spiegel.de | cio.de | tecchan nel.de | motor-talk.de |
|---|---|---|---|---|---|---|
| User-based CF | 0.28 | 0.28 | 0.23 | 0.26 | 0.17 | 0.15 |
| Mst. ppl. sequence | 0.23 | 0.32 | 0.31 | 0.15 | 0.10 | 0.16 |
| Item-based CF | 0.20 | 0.28 | 0.17 | 0.18 | 0.11 | 0.05 |
| Content-based | 0.07 | 0.15 | 0.09 | 0.13 | 0.07 | 0.31 |
| Most popular | 0.01 | 0.25 | 0.01 | 0.56 | 0.38 | 0.00 |

Website

*Figure 7: Recommendation Effectiveness on Different News Websites*

### 3.7 Recommendation frameworks

There are various recommendation frameworks such as Apache Lucene, Apache Mahout, LensKit (Ekstrand, Ludwig, Kolb, & Riedl, 2011), and MyMediaLite (Gantner, Rendle, Freudenthaler, & Schmidt-Thieme, 2011). Many of the frameworks focus on collaborative filtering. For content-based filtering, Apache Lucene is probably the "de-facto" standard. We use Apache Lucene for the recommender systems of both Docear and Mr. DLib.

### 3.8 Pre-calculating recommendations, or on-the-fly?

When developing a recommender system, one design decision is whether to calculate recommendations on the fly or in advance. Calculating one set of recommendations for a user in Docear took 52 seconds on average, with a standard deviation of 118 seconds (Beel, Langer, Gipp, et al., 2014). We therefore decided to pre-calculate recommendations. At certain events – e.g. when users added new papers to their mind-maps – the recommender system calculated a set of recommendations. When users requested recommendations, or Docear automatically showed recommendations to users, these pre-calculated recommendations were delivered. Mr. DLib calculates recommendations on-the-fly, which is possible because calculating document relatedness is less resource intensive than building sophisticated user models and analyzing dozens of mind-maps for each recommendation.

### 3.9 The randomization engine

To build an effective recommender system, A/B testing is essential. This means, the recommender system needs a pool of recommendation algorithms to choose from, and a logging mechanism to record, which algorithm was used and what the user feedback was (e.g. how many recommendations were clicked or downloaded). Unfortunately, in most situations, a simple A/B test with only two alternative algorithms will not be enough. In Mr. DLib, we implemented a "randomization engine" that first picks from several recommendation classes randomly, and then varies the parameters of the recommendation algorithms. The recommendation classes are content-based filtering (90% chance), stereotyping (4.9% chance), most-popular recommendations (4.9% chance), and random recommendations as baseline (0.2% chance). For each of the recommendation classes, the randomization engine chooses some parameter randomly. For instance, when content-based filtering is chosen, the system randomly selects whether to use "normal" terms, or key-phrases[13]. When key-phrases are chosen, the system randomly selects if key-phrases from the abstract, title, or title *and* abstract are used. Then the system randomly selects if unigrams, bigrams, or trigrams are used. Then the system randomly selects if one, two, three, … or twenty key-phrases are used to calculate document similarity.

Once the recommendations are calculated, the randomization engine chooses randomly if the recommendation candidates should be re-ranked based on bibliometrics. Re-ranking means that from the top *x* recommendations those ten documents are eventually recommended that have, for instance, the highest bibliometric score. Again, there are many parameters that are randomly chosen by the randomization engine. The engine selects the bibliometric (plain readership count, readership count normalized by age of the document, readership count normalized by the number of authors, etc.), the number of recommendation candidates to re-rank (10 to 100), and how the bibliometric and text relevance scores should be combined (bibliometric only, multiplication of scores, and some more variations). While we currently only work with readership data from Mendeley, we plan to obtain additional data from sources such as Google Scholar, Scopus, or Microsoft Academic (Sinha et al., 2015).

Developing such a randomization engine is a non-trivial task and we are not perfectly satisfied with our solution. It occurs too often that a fallback algorithm must be used because the randomly assembled algorithm cannot be applied because, for instance, a document does not have 20 key-phrases in its title.

### 3.10 Open source vs. closed-source

There are good reasons for both publishing software as open source and closed source (Comino, Manenti, & others, 2003; Mishra, Prasad, & Raghunathan, 2002; Paulson, Succi, & Eberlein, 2004; Raghunathan, Prasad, Mishra, & Chang, 2005). We decided to publish Mr. DLib' s source code as open source for the following reasons. First, we believe in the philosophy of open-source in general. Second, we do not consider the source code to be the valuable entity in the Mr. DLib project but the knowledge and partnerships. Third, in our experience, open-source projects are more likely to attract external volunteers and students, which is important because Mr. DLib is a small non-profit project and we cannot just hire software engineers, but require students and volunteers to participate.

[13] Keyphrases are the most meaningful phrases describing a document. Keyphrases are extracted with stemming, part-of-speech tagging and other mechanisms that we describe in an upcoming paper.

### 3.11 Copyright issues

We participated in other projects, were copyright was not clearly clarified, or developers only agreed to one particular license. In one case, the project owners wanted to change the project's license from one open-source license to another one. They had to contact everyone who had ever contributed to the project. It took several months before the license eventually could be changed. To prevent such a situation at Mr. DLib, everyone who joins the project as a developer agrees to the following.

> *[...] The general idea is that whatever you contribute to Mr. DLib can be used in the future by both you and the operator of Mr. DLib, in whatever way you or the operator want. This means: You keep the full rights on your own contributions to use them however you like. For instance, you may use your own contributions in any other project, publish the contribution wherever you want, under which license you want. [...] You grant the non-exclusive royalty-free permanent world-wide right to the operator to use your contribution in any way the operator wants. This includes, for instance, that the operator modifies your contributions, removes your contribution from the Mr. DLib project, sells your contribution, changes the license of your contribution, and uses the contribution in other projects. [...]*

Such an agreement gives the highest possible flexibility to the project owners, and ensures that Mr. DLib can change its license when necessary.

## 4 Recommender-System Operation

### 4.1 The need to deliver only good recommendations

In the recommender-system community, it is often reported that a recommender system should try to avoid making ‘bad’ recommendations as this hurts the users’ trust. In other words, it is better to recommend 5 good and 5 mediocre items than recommending 9 excellent but 1 bad item. To avoid bad recommendations, some relevance score is needed that indicates how good a recommendation is. Ideally, only recommendations above a certain threshold would then be recommended. However, at least Lucene has no such threshold that would allow a prediction of how relevant a recommendation is[14]. The Lucene text relevance score only allows to rank recommendations for one given query and compare the relevance of the results returned for that one query. In a recent analysis, we found that Lucene relevance scores and CTR correlate (Figure 8), but still it is not possible to avoid “bad” recommendations (Langer & Beel, 2017).

[14] Using the standard More-Like-This or search function https://wiki.apache.org/lucene-java/ScoresAsPercentages

Even if Lucene had an “absolute” text relevance score, this score would only be able to prevent bad recommendations to some extent. We see a high potential in bibliometrics to support recommender systems in not recommending bad items.

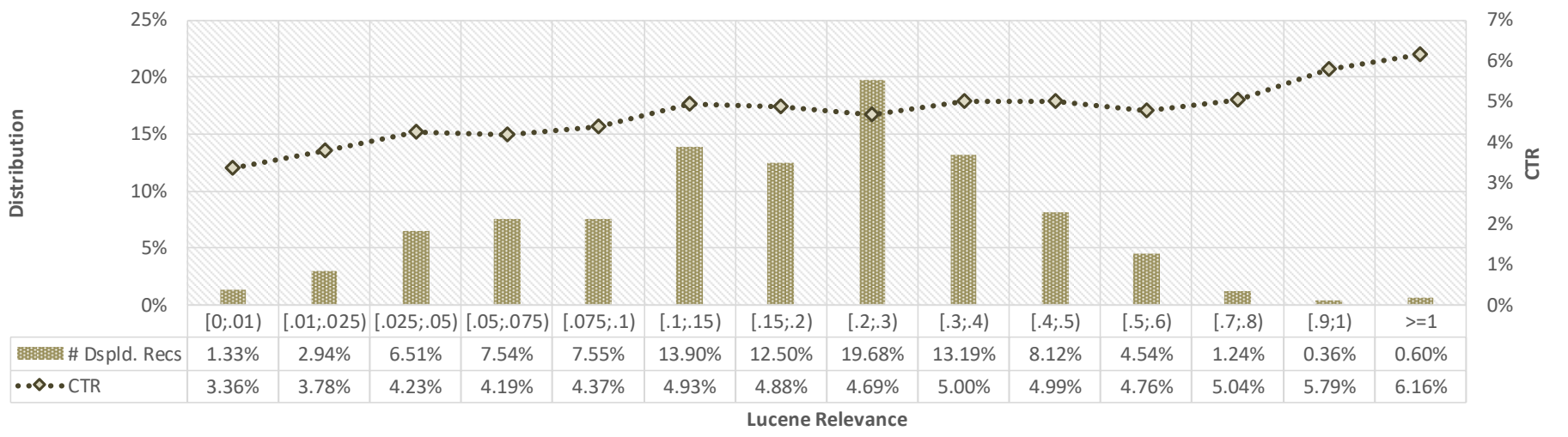

*Figure 8: Lucene relevance scores and CTR* (Langer & Beel, 2017)

## 4.2 Ensure timely delivery of recommendations

On the Internet, users tend to be impatient: the longer they wait for content, the less satisfied they become (Guse, Schuck, Hohlfeld, Raake, & Möller, 2015; Kim, Xiong, & Liang, 2017; Selvidge, Chaparro, & Bender, 2002). This holds true for recommender systems, too. We observed that the longer users had to wait for recommendations, the less likely they were to click a recommendation. Figure 9 shows that processing time[15] for most recommendations (34%) was between 1 and 2 seconds. These recommendations also had the highest click-through rate (CTR) of 0.15% on average. In contrast, recommendations that needed 7 to 8 seconds for calculation had a CTR of 0.08%.

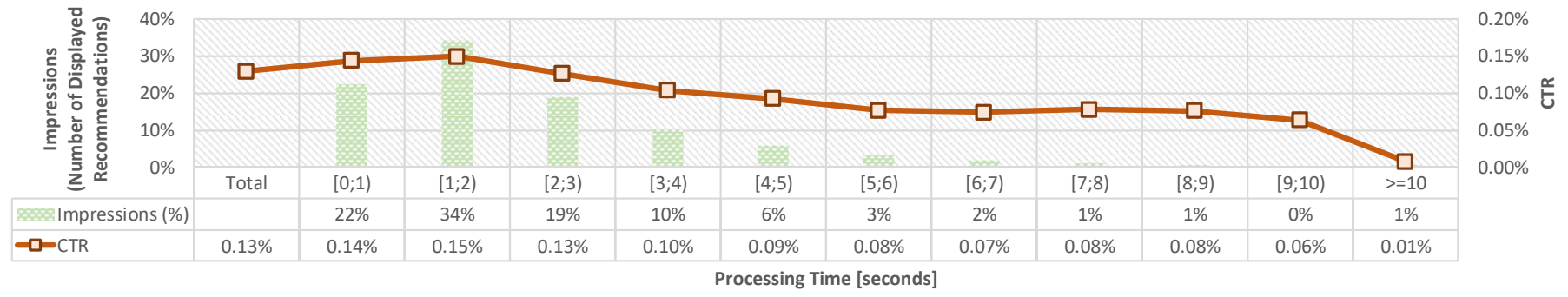

*Figure 9: Click-Through Rate (CTR) based on Processing Time*

The re-ranking of recommendations based on bibliometrics requires rather a lot of calculation (primarily due to the randomization engine and because we store many statistics when calculating the bibliometrics). Consequently, when evaluating the effectiveness of bibliometric re-ranking, one need to additionally consider if the additional effectiveness if worth the additional time users need to wait.

[15] “Processing Time” is the time from receiving a request until delivering the response on side of Mr. DLib. It may be that a user who must wait too long, leaves the web page and does not even see the recommendations.

### 4.3 Number of recommendations

Another question that may seem simple to answer is how many recommendations to display? One? Two? ...Ten? We experimented in Mr. DLib with varying numbers of recommendations between 1 and 15 (Beierle, Aizawa, & Beel, 2017). We observed that the more recommendations were displayed, the lower click-through rate became (Figure 10). From these results, one cannot conclude how many recommendations to display, and more research is necessary.

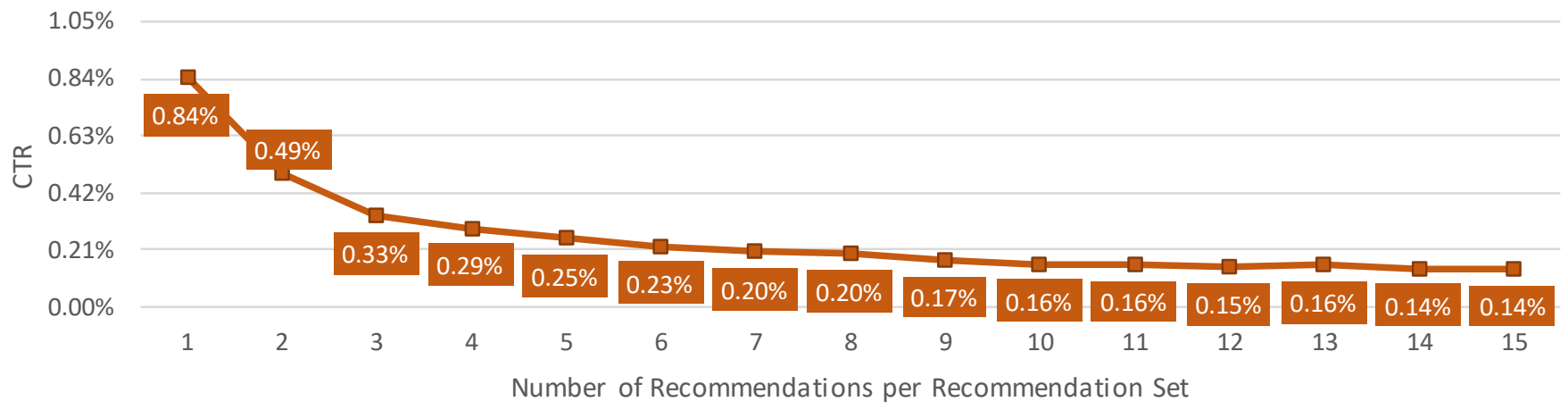

*Figure 10: Number of displayed recommendations and CTR* (Beierle et al., 2017)

### 4.4 Persistence

Another unanswered question is if, when, and how often to re-display recommendations for documents that were recommended previously to a user. We observed that recommendations being re-shown to a user typically received lower click-through rates than 'fresh' recommendations (Beel, Langer, Genzmehr, & Nürnberger, 2013b). For instance, if a recommendation was shown the third time to a user, CTR was 6.97%, while “fresh” recommendations being shown for the first time received a CTR of 9.59% (Figure 11).

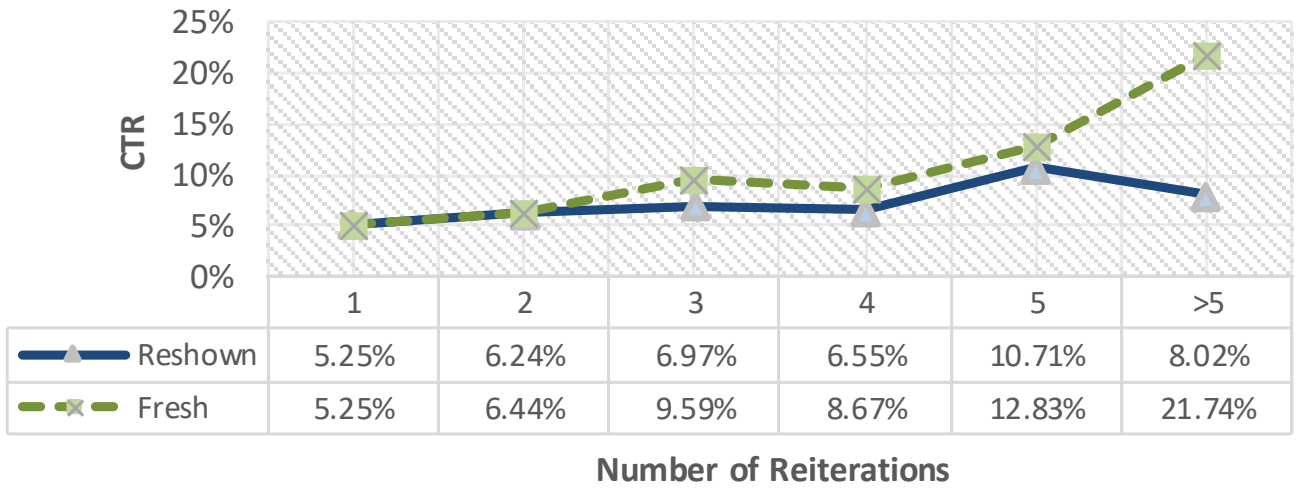

*Figure 11: Fresh vs. Re-shown recommendations*

However, when looking at re-shown recommendations with a 1-day delay, the picture changes (Figure 12). In this analysis, we ignored recommendations that were reshown to the same user within 24 hours. In this case, CTR for both fresh and reshown recommendations was around the same. For instance, when recommendations were shown the second time, CTR was 7.72%. Fresh recommendations had a CTR of 6.69% (the difference is statistically not significant). Again, no clear conclusions can be drawn from this results.

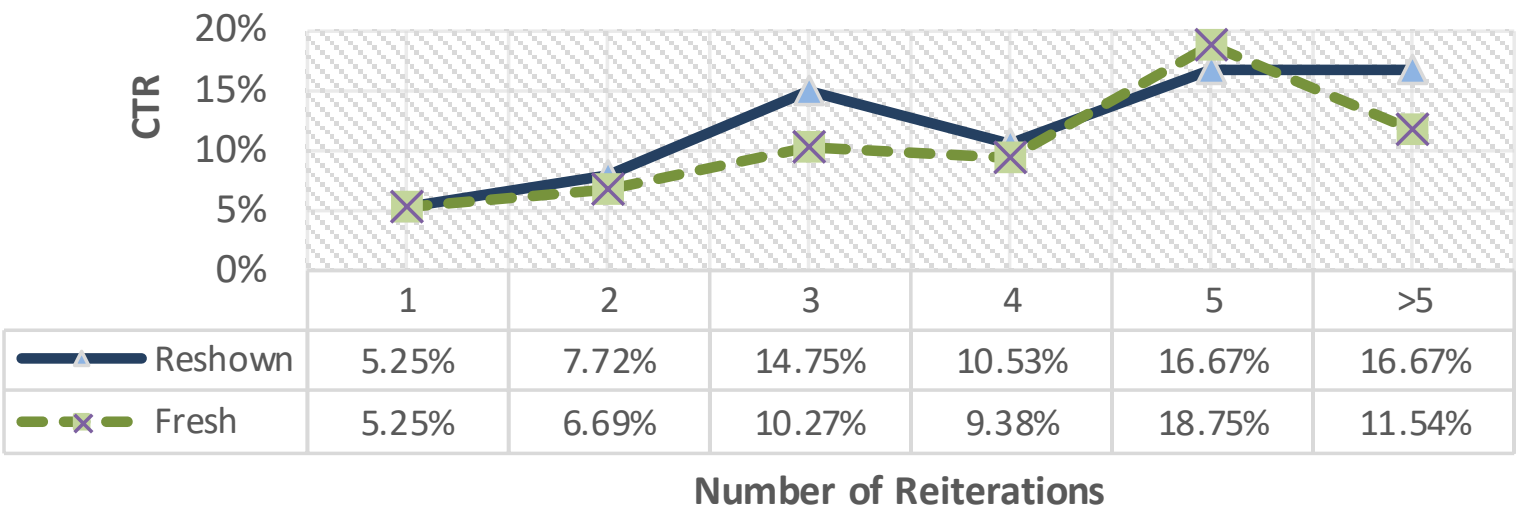

*Figure 12: Fresh vs. Re-shown recommendations (1-day delay)*

### 4.5 Monitoring

When running a production system, uptime is crucial. To be informed and keep Mr. DLib's partners informed about our servers' uptime, we currently use Uptimerobot.com, a free monitoring tool. The monitoring data is publicly available at http://monitoring.mr-dlib.org (Figure 13) and the tool checks in 5-minute intervals, if Mr. DLib's API is available. The tool also records the response time (Figure 14).

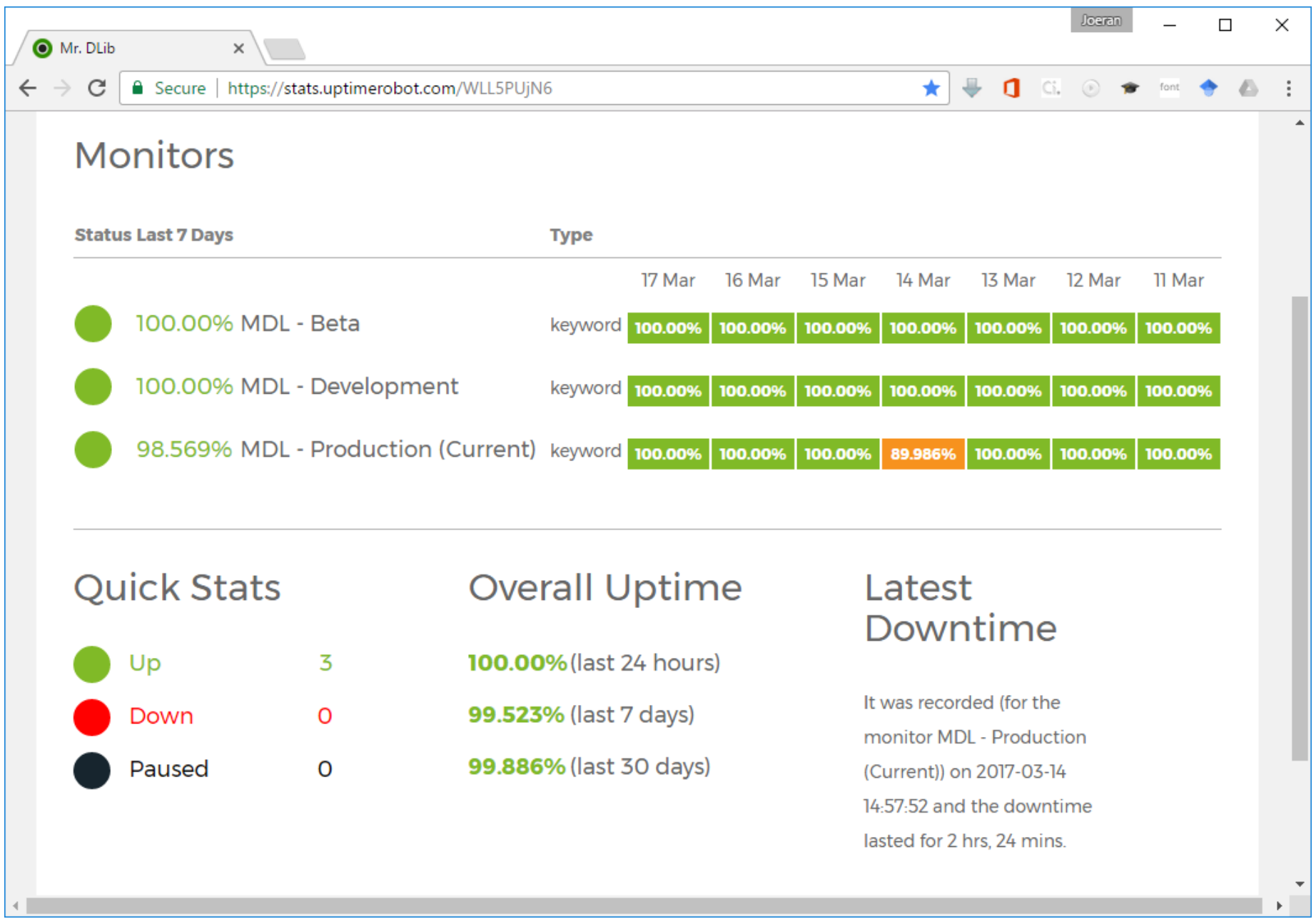

*Figure 13: Screenshot of the Monitoring Dashboard (Uptime)*

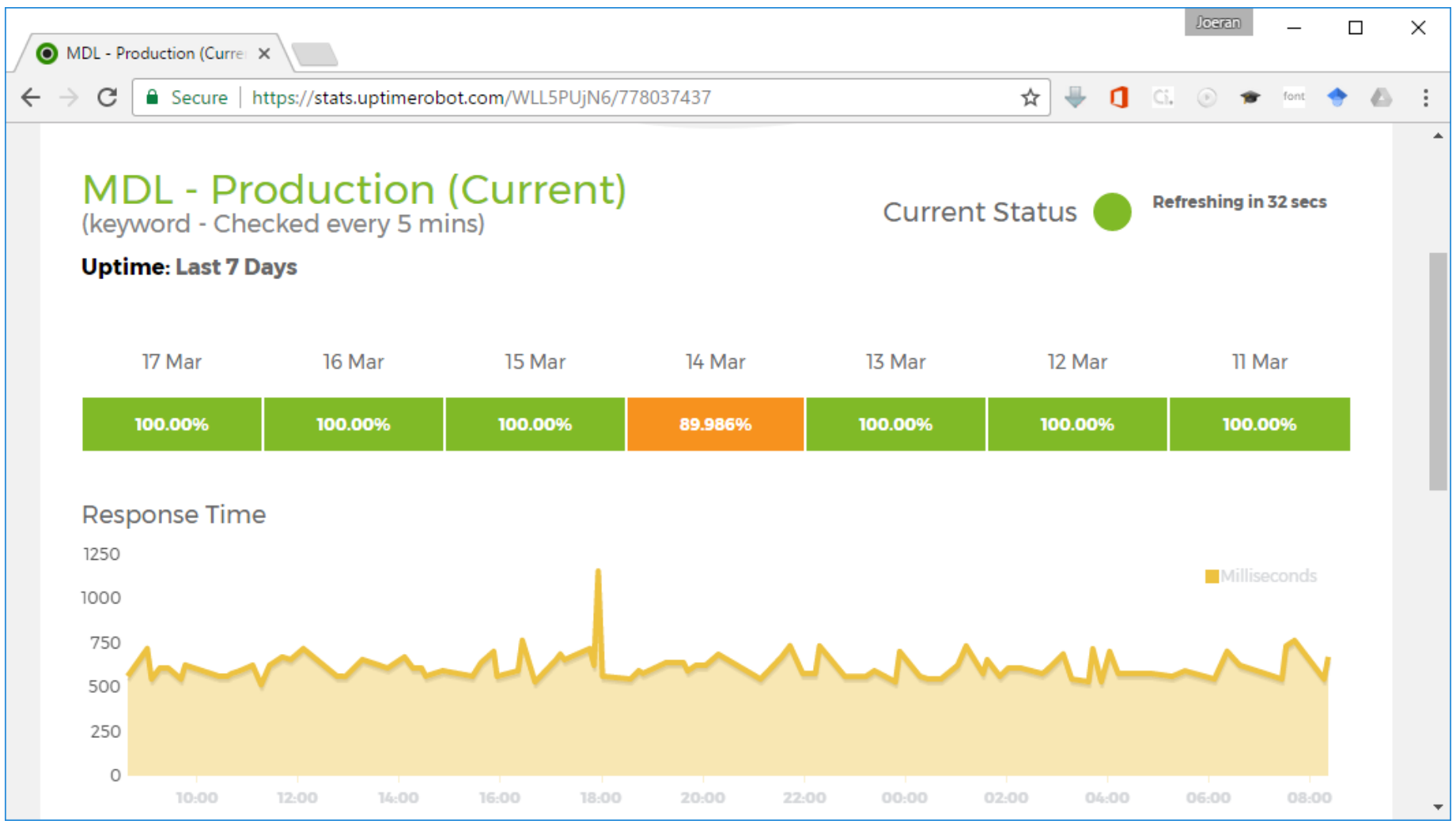

*Figure 14: Screenshot of the Monitoring Dashboard (Response Time)*

## 4.6 Data quality

For good content-based recommendations the quality of metadata is crucial. When using publicly available datasets, the quality is often rather good because the datasets were designed and maybe even curated for that purpose. In the real-world, however, data quality is often low. For instance, some of the most productive "authors" in Mr. DLib's database are "et al.", "and others", and "AnoN" (Table 1). This is a problem we share with many other operators of academic services. For instance, one of the most cited authors on Google Scholar is "et al.". When this data is used e.g. to recommend papers of co-authors, or to re-rank recommendations based on h-index, the resulting recommendations will be of suboptimal quality.

*Table 1: Some of the most productive "authors"*

| Author | Document Count |
|---|---|
| et al. | 35,865 |
| and others | 26,331 |
| AnoN. | 25,719 |
| Anonymous | 21,933 |
| [Unknown] | 17,210 |
| [[author]]??? | 17,191 |
| Unknown | 16,233 |
| u.a. | 16,162 |

Cleaning data from third parties is a labor-intensive and usually boring task (it is much easier to motivate a colleague to implement a novel recommendation algorithm than convincing a colleague to spend some weeks cleaning data in a database). Data cleaning becomes particularly challenging, when the data comes from a third party and the data are updated occasionally by the partner. In that case, one would need a process to decide how to deal with the updated data and judge if the new data from the partner is better than the manually changed data in our system. From our experience in other projects, we know that manually cleaning data usually causes many problems. Therefore, we decided to do no manual data cleaning in Mr. DLib, and only apply a few heuristics such as ignoring “et al.” when calculating bibliometrics.

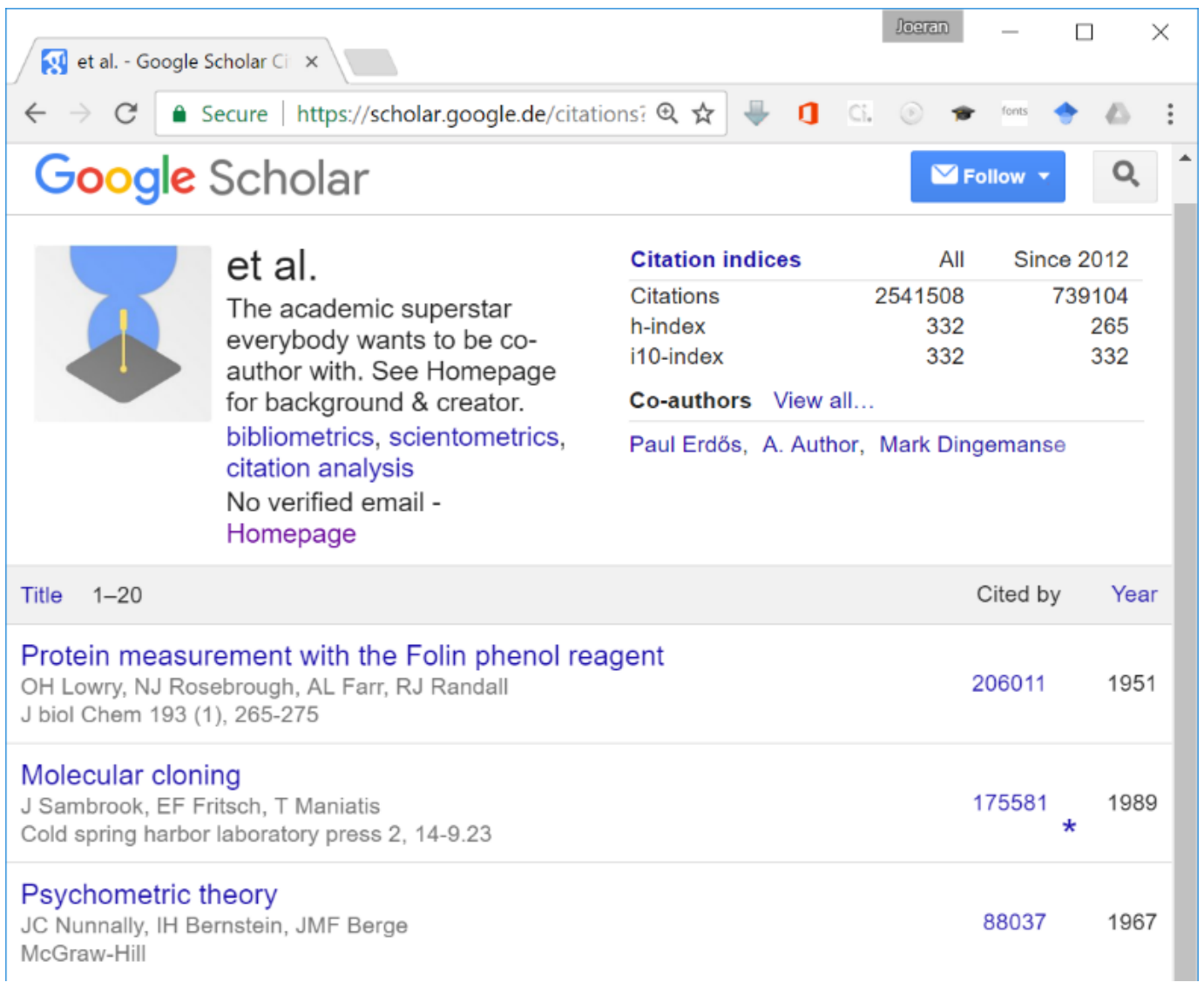

*Figure 15: The author's profile of "et al." on Google Scholar*

## 4.7 Cleaning the recommendation list (remove duplicates and near-duplicates)

In Mr. DLib's first version, the recommendation list often contained (near-)duplicates (Figure 16). The reason was that Sowiport had duplicate documents in its corpus, sometimes with minor variations such as different spellings of author names (e.g. “J. Smith” vs. “John Smith”). We implemented a simple mechanism that removed duplicates from the recommendation list based on a “clean-title” filtering.

The clean title equals the original title but all special characters and numbers are removed. We know from other operators of academic recommender systems that they use similar methods. For instance, one operator considers documents to be duplicates when clean title and publication year are the same. To the best of our knowledge, there is no research on how to remove duplicates most effectively from recommendation lists.

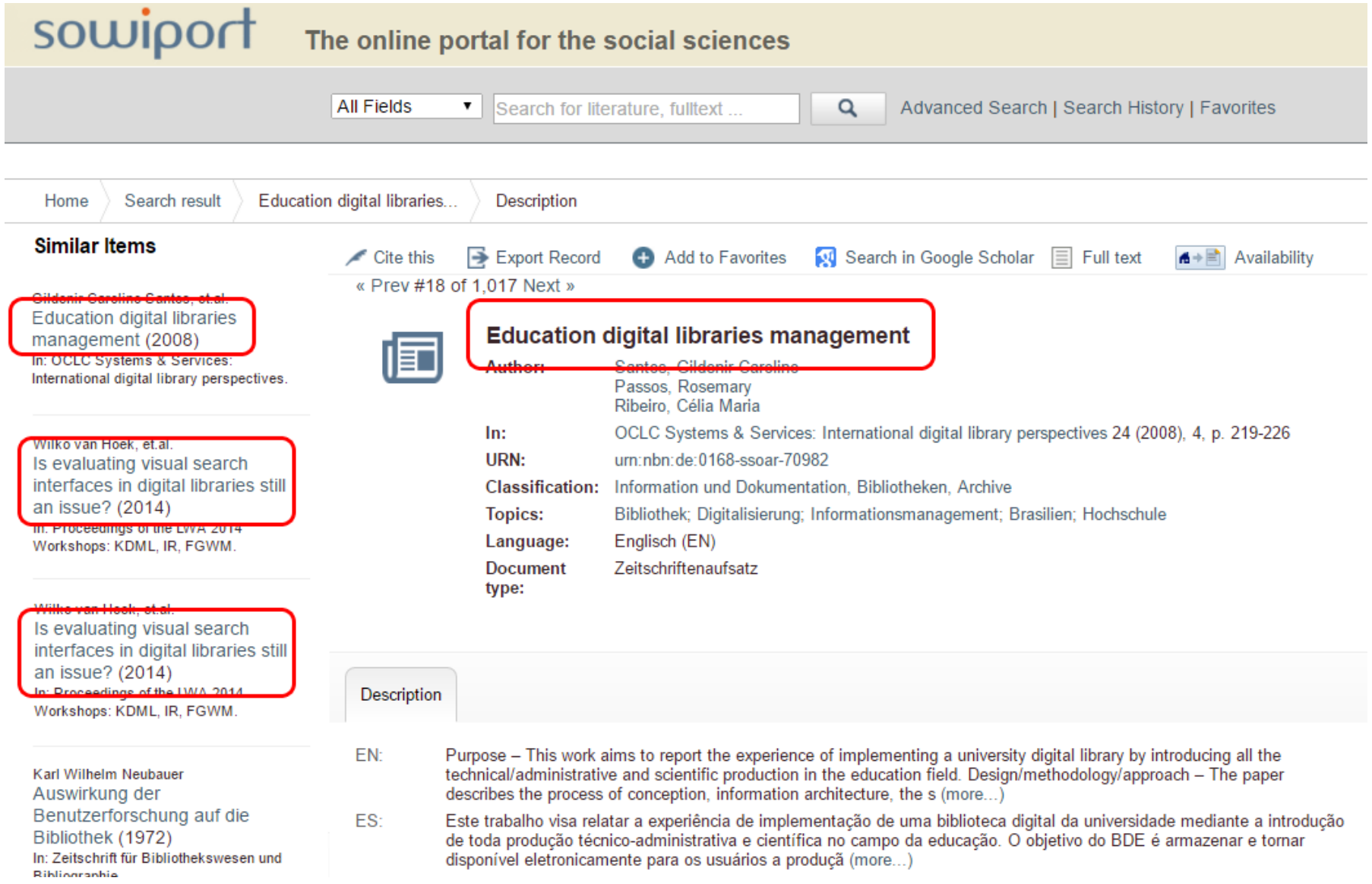

*Figure 16: Duplicate entries in the recommendation list*

## 4.8 Dealing with Web Crawlers

When Sowiport first integrated Mr. DLib, the recommendations were integrated on Sowiport's servers. This means, once a user visited a web page on sowiport.de, the Sowiport servers built the web page, requested recommendations, added the recommendations to the web page, and then delivered the web page to the user. This had the disadvantage that also robots crawling the Web received recommendations, which resulted in a high number of recommendations delivered (around 2.5 million recommendation sets per month[16]). However, the logging of clicks was realized with JavaScript, which is usually not executed by web crawlers. Hence, we had many recommendations but only few clicks.

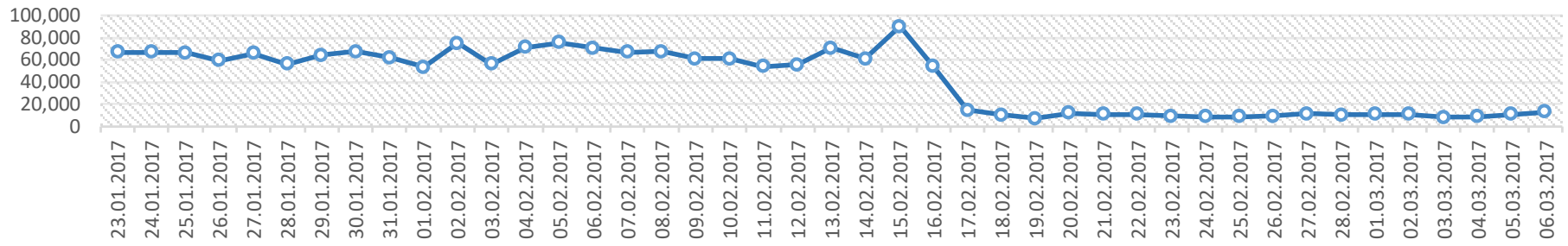

*Figure 17:Number of requests before and with JavaScript Client*

Since mid of February 2017, Sowiport uses a JavaScript to request recommendations. This JavaScript is loaded once the normal web page is delivered to a user, and web crawlers typically do not execute the JavaScript, hence do not receive recommendations. Since the use of the JavaScript, the number of requested recommendations strongly decreased. Before the JavaScript was used, 60 thousand recommendations were requested on average per day. With the JavaScript, around 10 thousand recommendation sets are requested per day on average.

[16] One set of recommendation contains between one and fifteen recommendations

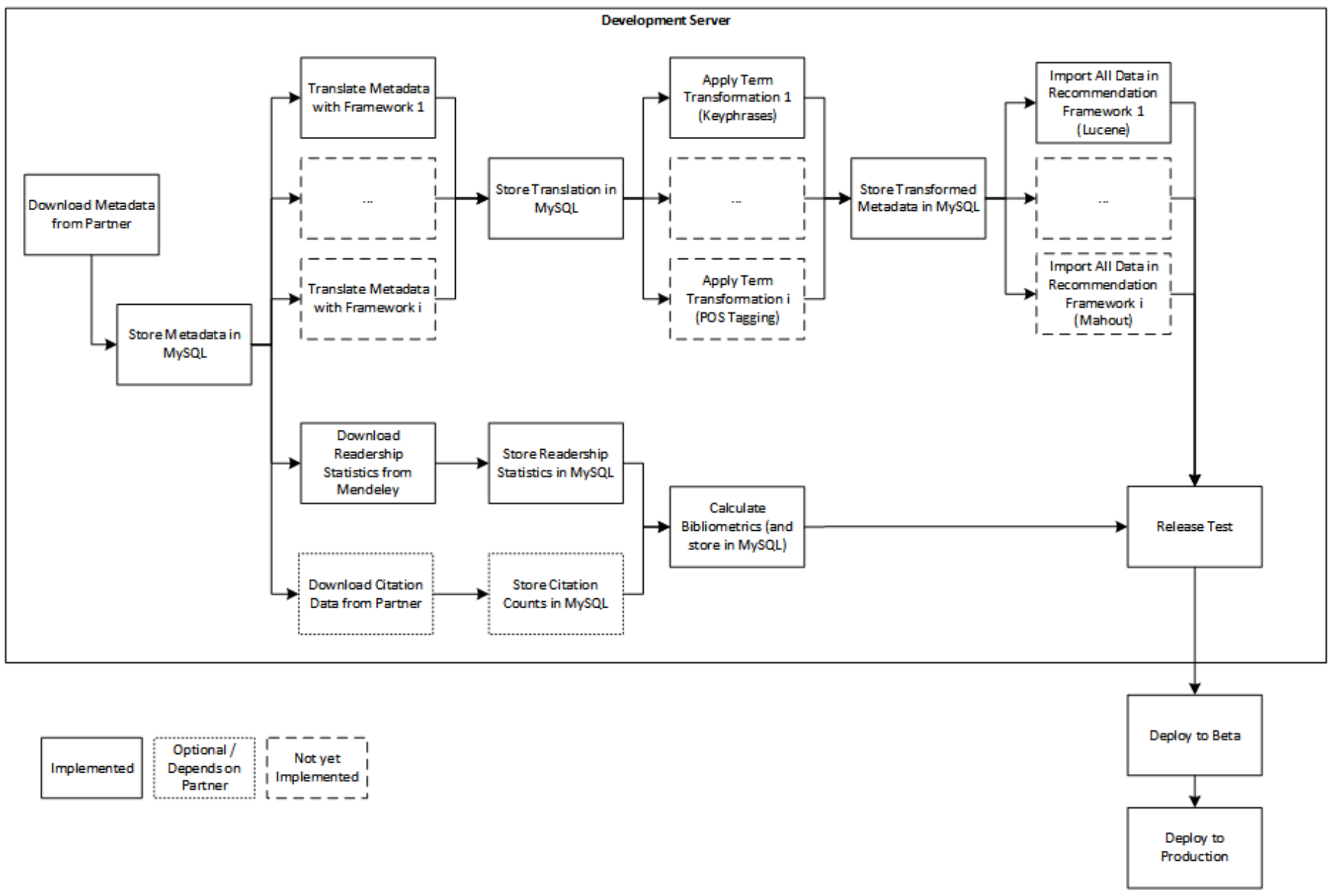

*Figure 18: Adding New Content*

### 4.9 Managing many processes

A recommender system may seem like a rather simple system that only needs to import some data, calculate recommendations, and maybe do some logging. However, a recommender system quickly becomes a complex system, especially when dealing with different data sources and recommendation approaches. Figure 18 shows our process for adding new content from a partner. The process includes parsing the partners export (e.g. XML) and storing the data in our MySQL database, translating metadata for multilingual content-based filtering, requesting readership statistics from Mendeley, calculating bibliometrics, indexing everything in Lucene (and other recommendation frameworks soon), and then deploying and testing on the development, beta, and production server. The process is mostly sequential, which means that adding metadata from a partner for the first time may take several weeks.

## 5 Recommender-System Research

### 5.1 Expect CTR to decrease over time

On both Docear and Mr. DLib we observed that CTR decreases over time, although in the very last week of the data analysis period CTR increased again (Figure 19). We do not know the reasons but will analyze the data in more detail soon. A decreasing CTR is not ideal as it indicates that the relevance of recommendations decreases over time. In addition, less clicks mean that it becomes more difficult to obtain statistically significant results. We are not aware of many other recommender-system operators who report on CTR over time. However, some do. Middleton, Shadbolt, & De Roure (2004) observed a similar effect as we did, i.e. CTR in their recommender system decreased over time. Jack (2012b) reports the opposite, namely that precision increased over time (p=0.025 in the beginning, p=0.4 after six months).

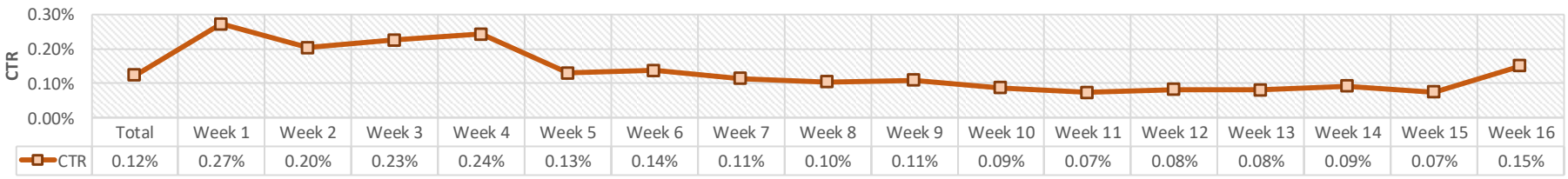

*Figure 19: CTR of Mr. DLib between 2016-10-18 and 2017-01-31 (45 million recommendations overall)*

## 5.2 Apparently implausible data

Sometimes, data seems not plausible. For instance, in Mr. DLib, 4.7% of all clicks occur more than five days after the recommendations were delivered (Figure 20). We consider it also rather unlikely that only 4.8% of clicks occur within the first 30 seconds after recommendations were delivered[17]. We have not yet analyzed this in more detail. However, a deeper analysis requires changes on our logging methods, and maybe also changes on the side of our partner, and we have not yet had the time for this. Situations like this happen occasionally and require often a rather high effort to analyze and fix.

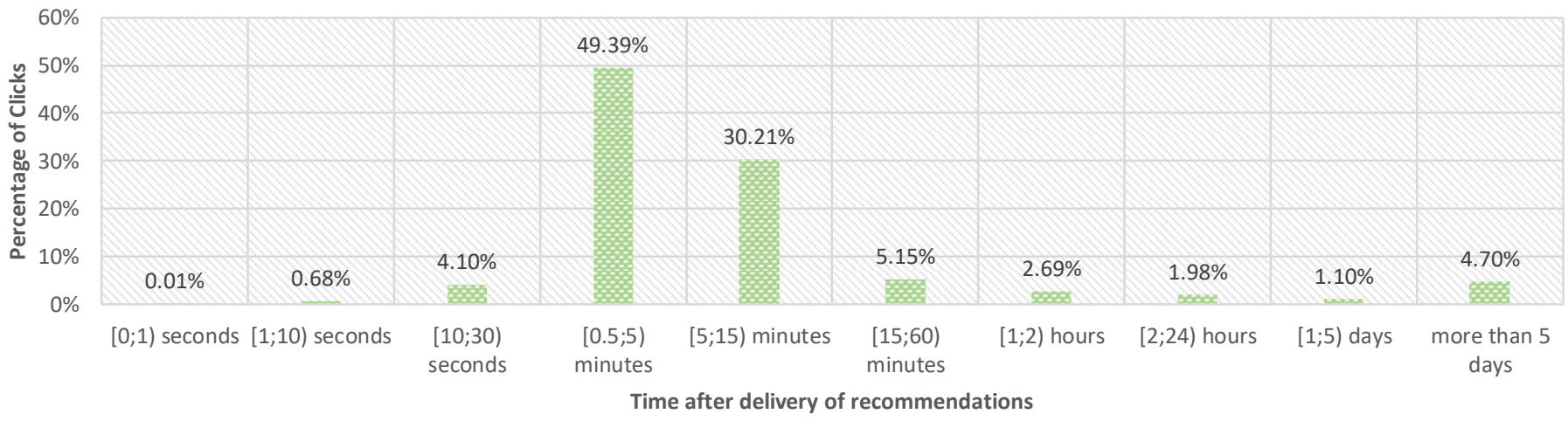

*Figure 20: Time after which recommendations were clicked*

## 5.3 No tweaking of data

In offline evaluations, it is common to tweak datasets. For instance, Caragea et al. removed papers with fewer than ten and more than 100 citations from the evaluation corpus, as well as papers citing fewer than 15 and more than 50 papers (Caragea, Silvescu, Mitra, & Giles, 2013). From originally 1.3 million papers in the corpus, around 16,000 remained (1.2%). Pennock et al. removed documents with fewer than 15 implicit ratings from the corpus (Pennock, Horvitz, Lawrence, & Giles, 2000). From originally 270,000 documents, 1,575 remained (0.58%). Such tweaking and pruning of datasets may be convenient for the research and lead potentially to high precision. However, applying a recommendation approach to only 0.58% of the documents in a corpus, will lead to a very poor recall, i.e. the results that have little relevance for running a real-world recommender system. In other words, in a real-world recommender system such a tweaking would be difficult, unless one would accept that recommendations can be delivered only for a fraction of documents in the corpus.

[17] Please note that we record the time when Mr. DLib delivers recommendations to the partner, and the time when a user clicks a recommendation. Between delivery to the partner and displaying recommendations on the website, some time passes (probably one or two seconds).

### 5.4 Accept Failure

When we reviewed over 200 research articles about recommender systems, every article that introduced a new recommendation approach reported to outperform the state-of-the art (Beel, Gipp, et al., 2016). We were not that lucky.

We re-ranked content-based recommendation with readership data from Mendeley (Siebert et al., 2017), but results were not as good as expected. We used a key-phrase-extraction approach to improve our content-based filtering approach, and experimented with a variation of parameters: we varied the text-fields from which key-phrases were extracted, we varied the number of key-phrases being used, and we varied the type of key-phrases (unigrams, bigrams, trigrams, or a mix). None of these variations performed better than the simple out-of-the-box Apache Lucene baseline.[18] We experimented with stereotype and most-popular recommendations, two approaches that are effective in domains such as movie recommendations and hotel search (Beel, Dinesh, et al., 2017). Again, the approaches performed not better than Apache Lucene.

There are a few potential reasons why our experiments delivered disappointing results (besides the possibility that the recommendation approaches are just not effective). For instance, in case of the bibliometric re-ranking, we assume that click-through rate might be not appropriate to measure the re-ranking effectiveness. However, even while there might be plausible reasons for the results, the fact remains that many of our experiments delivered rather disappointing results, but this is probably rather the rule than the exception in a real-world scenario.

### 5.5 Other researchers' interest in datasets

One advantage of working on a real-world recommender system is the possibility to release datasets, which then can be used (and cited) by other researchers. Some datasets such as MovieLens are very popular in the recommender system community. The MovieLens dataset was downloaded 140,000 times in 2014 (Harper & Konstan, 2016), and Google Scholar lists 10,600 papers that mention the MovieLens dataset (Figure 21)[19].

[18] The analysis is still in process, and the final results will be published soon.

[19] https://scholar.google.com/scholar?hl=en&q="movielens"

Not all datasets become that popular. In September 2014, we released a dataset of Docear's recommender system (Beel, Langer, Gipp, et al., 2014). The datasets contained metadata of 9.4 million articles, including 1.8 million articles publicly available on the Web; the articles' citation network; anonymized information on 8,059 Docear users; information about the users' 52,202 mind-maps and personal libraries; and details on the 308,146 recommendations that the recommender system delivered. In the 2.5 years since publication, 31 researchers requested to download the dataset. To the best of our knowledge, none of these researchers eventually has analyzed the dataset and published their results.

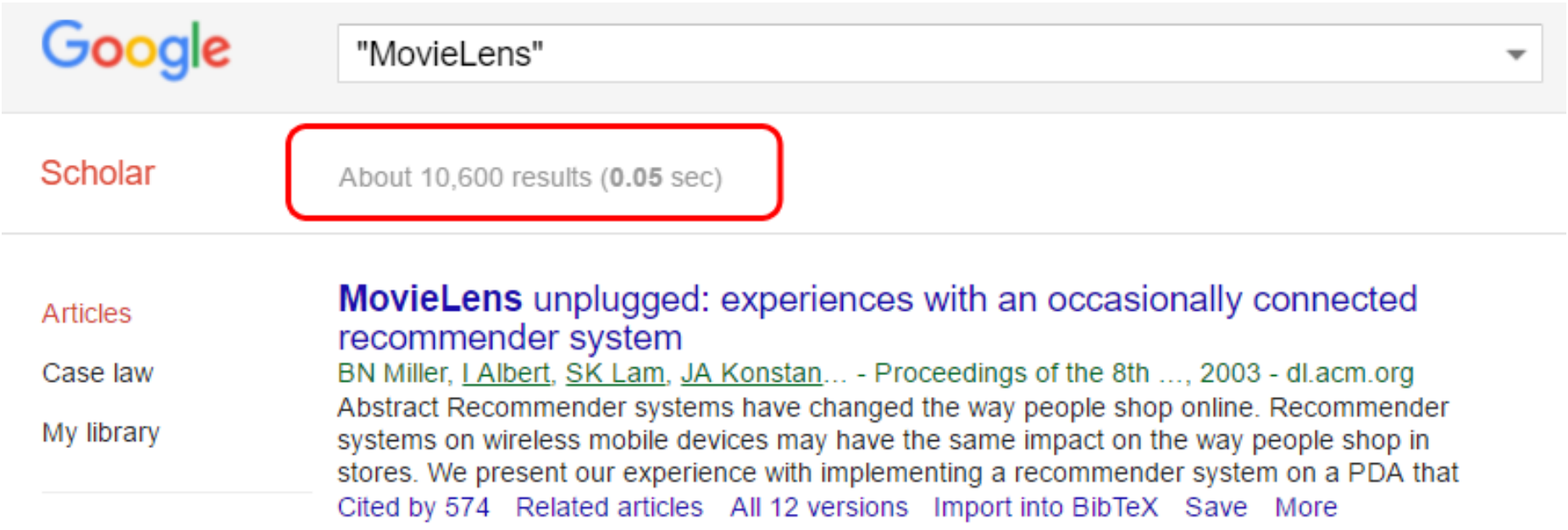

*Figure 21: Number of search results on Google Scholar for "MovieLens"*

## 6 Summary & Conclusion

Building and operating real-world recommender systems is a time-consuming and challenging task, as is research on such systems.

To develop recommender systems, knowledge from various disciplines is required such as machine learning, server administration, databases, web technologies, and data formats. When building our own recommender systems, we could find no guidance in the literature. Most published research results had questionable evaluations and even if evaluations were sound, recommender systems seem to perform just too differently in different scenarios. Consequently, we used Apache Lucene as recommendation framework and began from scratch. Evaluating different recommendation approaches requires a randomization engine that selects and assembles recommendation algorithms automatically. In addition, a detailed logging mechanism is required that records which algorithm created which recommendation and how the user reaction was to the different algorithms. Another challenge lies in dealing with sub-optimal data quality from partners. Due to time constraints, we decided to not manually improve the data but just work with what we got.

Running a recommender system requires the fast generation and delivery of recommendations. Otherwise, users become dissatisfied and click-through rates decrease. Although it is widely known that recommender systems should avoid making bad recommendations, this is not easily accomplished in practice, at least if Lucene is used as recommendation framework. Lucene has no absolute relevance score and hence no mechanism such as "recommend only items above a certain relevance threshold" can be implemented.

Research on real-world recommender systems is probably more frustrating than research in a lab-environment, mostly because data cannot be tweaked that easily. Consequently, we had to accept that many experiments with novel recommendation approaches failed. Similarly, while some recommendation datasets such as MovieLens are widely used in the community, we could not yet manage to establish our datasets as interesting source for other researchers.

Despite all these challenges, research on real-world recommender systems is a rewarding and worthwhile effort. We feel that working on real systems provides much more relevant research results. In addition, offering a real-world open-source project attracts many volunteers, students, and project partners. This, in turn, enabled us to conduct research in many areas, and be quite productive in terms of publication output.

## 7 Acknowledgements

This work was supported by a fellowship within the FITweltweit programme of the German Academic Exchange Service (DAAD). In addition, this publication has emanated from research conducted with the financial support of Science Foundation Ireland (SFI) under Grant Number 13/RC/2106. We are also grateful for the support received by Sophie Siebert, Stefan Feyer, Felix Beierle, Sara Mahmoud, Gabor Neusch, and Mr. DLib's partners.